\documentclass[notitlepage,aps,amsmath,amssymb,graphicx,onecolumn]{revtex4-1}

\usepackage{graphicx}
\usepackage{dcolumn}
\usepackage{bm}
\usepackage{float}
\usepackage[caption=false]{subfig}
\usepackage{amsmath}
\usepackage{amssymb}
\usepackage[utf8]{inputenc}
\usepackage[T1]{fontenc}
\usepackage{mathptmx}
\usepackage[top=30truemm,bottom=30truemm,left=25truemm,right=25truemm]{geometry}

\begin{document}

\title{Two-dimensional simulated tempering for the isobaric-isothermal ensemble with fast on-the-fly weight determination}

\author{Hiromune Wada$^{1}$\thanks{hwada@tb.phys.nagoya-u.ac.jp} and Yuko Okamoto$^{1,2,3}$\thanks{okamoto@tb.phys.nagoya-u.ac.jp}}

\affiliation{
$^{1}$Department of Physics, Graduate School of Science, Nagoya University, Nagoya, Aichi 464-8602, Japan \\
$^{2}$Center for Computational Science, Graduate School of Engineering, Nagoya University, Nagoya, Aichi 464-8603, Japan \\
$^{3}$Information Technology Center, Nagoya University, Nagoya, Aichi 464-8601, Japan
}




\pacs{Valid PACS appear here}


\begin{abstract}
We propose a method to extend the fast on-the-fly weight determination scheme for simulated tempering to two-dimensional space including not only temperature but also pressure.
During the simulated tempering simulation, weight parameters for temperature-update and pressure-update are self-updated independently according to the trapezoidal rule.
In order to test the effectiveness of the algorithm, we applied our proposed method to a peptide, chignolin, in explicit water.
After setting all weight parameters to zero, the weight parameters were quickly determined during the simulation.
The simulation realised a uniform random walk in the entire temperature-pressure space.
\end{abstract}


\maketitle

\section{Introduction}
In conventional molecular dynamics (MD) and Monte Carlo (MC) simulations of complex systems such as biomolecules, the sampling efficiency of the configuration space is reduced due to the ruggedness of the energy landscape of the system.
More specifically simulations at low temperatures get trapped in states of multiple local minima in energy landscapes.
To overcome this difficulty, generalised-ensemble algorithms were proposed~\cite{hansmann1999generalized,mitsutake2001generalized}.
One of the widely known ideas in generalised ensemble is referred to as \textit{Tempering}~\cite{lyubartsev1992new,marinari1992simulated,swendsen1986replica,geyer1991computing,hukushima1996exchange,sugita1999replica}.
In \textit{Tempering}, temperature is treated as a dynamical, discrete variable, and a random walk is realised in the temperature space while maintaining detailed balance conditions.
By random walking in the temperature space, simulations can escape from the local minima in energy landscapes of the system.
There are two well-known tempering methods. The first is the simulated tempering (ST) method~\cite{lyubartsev1992new,marinari1992simulated} that performs a random walk in the temperature space by giving an extra weight parameter, which is Helmholtz free energy.
The ST method and its improved versions have been studied by several groups and have been used for various problems in the field of molecular simulation such as spin systems and biomolecular systems~\cite{Irback1995,Hansmann1997,Irback1999,mitsutake2000replica,mitsutake2004,Kim2004,park2007choosing,park2008comparison,zhang2008comparison,Kim2010,mori2010generalized,nagai2012simulated,nguyen2013communication,zhang2015folding}.
The other is the parallel tempering method (also known as replica-exchange method: REM)~\cite{swendsen1986replica,geyer1991computing,hukushima1996exchange,sugita1999replica} in which non-interacting replicas of the system are prepared and temperatures are exchanged between the replicas to realise random walks in the temperature space without the need for weight parameters.
In comparative studies of ST and REM, it is concluded that ST has higher transition probabilities between different temperatures and has a higher rate of random walk diffusion in temperature space than REM~\cite{mitsutake2000replica,zhang2008comparison,park2008comparison}.
ST has better sampling efficiency than REM because the faster the temperature diffusion is, the faster is the diffusion rate in the potential energy space.
However, REM is more widely used than ST.
This is because while ST requires a tedious and difficult weight parameter determination process before the production simulation, REM does not require this process.
In recent years, the on-the-fly weight determination scheme~\cite{nguyen2013communication} has been proposed, and it has greatly simplified the weight determination process.
With this, we expect that ST will be more frequently used in the future.

ST has also been generalised to realise a random walk in a multidimensional space by adding, e.g., pressure~\cite{mori2010generalized} and an external magnetic field~\cite{nagai2012simulated} to temperature, which is referred of as the multidimensional ST~\cite{mitsutake2009multidimensional1,mitsutake2009multidimensional2,mitsutake2009simulated} (for a review, see, e.g., Ref.~\cite{mitsutake_mori_okamoto}).
However, because the on-the-fly weight determination scheme has yet to be extended to the multidimensional ST, the difficulty of weight parameter determinations remains.
It is desired to extend the on-the-fly weight determination scheme to the two-dimensional space such as temperature and pressure.
In this article, we propose a two-dimensional on-the-fly weight determination scheme for the two-dimensional ST in the isobaric-isothermal ensemble.

This article is organised as follows.
In section \ref{sec:method} we explain the methods.
Section \ref{sec:computational} gives the computational details.
Section \ref{sec:result} presents the results of applications of the present method.
Section \ref{sec:conclusion} is devoted to conclusions.


\section{methods}
\label{sec:method}

\subsection{Simulated tempering}
We first review the ST algorithm.
In this method, temperature itself becomes a dynamical variable which takes discrete values $T_n\:(T_1<T_2<...<T_N)$.
Distribution function of a state at temperature $T_n$ and potential energy $E$ are given by the following formula:
\begin{align}
W_\mathrm{ST}(E;T_n)=\mathrm{e}^{-{\beta}_n E+f_n},
\end{align}
where $\beta_{n}=1/k_\mathrm{B}T_n$ ($k_\mathrm{B}$ is the Boltzmann constant).The function $f_n=f(T_n)$ is chosen so that the distribution function $p_\mathrm{ST}(T_n)$ of temperature may be uniform:
\begin{align}
p_\mathrm{ST}(T_n) &=\int dEn(E)W_\mathrm{ST}(E;T_n)\notag\\
&=\int dEn(E)\mathrm{e}^{-\beta_nE+f_n}\notag\\
&\equiv\mathrm{const.} \label{conv_st}
\end{align}
Here, $n(E)$ is the density of states.
From Eq.~\eqref{conv_st}, the function $f_n$ is given by
\begin{equation}
\label{helmholtz}
f_n = -\ln\int dEn(E)\mathrm{e}^{-\beta_nE}.
\end{equation}
Hence, $f_n$ are dimensionless Helmholtz free energy at temperature $T_n$.

Once the parameters $f_n$ are determined, a ST simulation is realised by repeating the following two steps:
\begin{enumerate}
\item Perform a canonical MC or MD simulation for a certain number of steps at temperature $T_n$.
\item Update the temperature $T_n$ to the neighbouring value $T_{n\pm 1}$ while fixing the configurations. The transition probability of this process is given by the following Metropolis criterion~\cite{metropolis1953equation}:
\begin{equation}
\label{st_met}
w(T_n,T_{n\pm1})=\min(1,\exp(-\Delta)),
\end{equation}
where
\begin{equation}
\label{st_delta}
\Delta=(\beta_{n\pm1}-\beta_{n})E-(f_n-f_{n\pm1}).
\end{equation}
\end{enumerate}


\subsection{On-the-fly estimation of weight parameters}
In the ST weight determination process, it is a common strategy to perform short trial simulations and estimate the weights from multiple-histogram reweighting technique~\cite{kumar1992weighted,shirts2008statistically,mitsutake2000replica}.
However, this method of assigning weight parameters has a risk that the exact weight parameters cannot be obtained when the trial simulation is too short.
If weight parameters with insufficient accuracy are used, a sufficient random walk in the temperature space cannot be performed, resulting in a poor ST performance.
As a solution to this problem, Nguyen \textit{et al.} proposed on-the-fly weight determination scheme that dynamically updates the weight parameters during simulation~\cite{nguyen2013communication,zhang2015folding}.
This method does not require any prior trials.

This method is based on the following formula which gives a near optimal weight parameter proposed by Park and Pande~\cite{park2007choosing}:
\begin{align}
\label{park_pande}
f_{n+1}=f_{n} + ({\beta}_{n+1}-{\beta}_{n})\frac{\bar{E}_{n+1}+\bar{E}_{n}}{2},
\end{align}
where $\bar{E}_{n}$ is the average potential energy at temperature $T_n$.

The protocol for determining the weight parameters is as follows:

\begin{itemize}
\setlength{\parskip}{0cm}
\setlength{\itemsep}{0cm} 
\item First, set weight parameters $f_{n}=0$.
\item Start the ST simulation at the lowest temperature $T_1$.
Accumulating potential energy and calculating the average potential energy $\bar{E}_{1}$,
we obtain the weight parameter $f_2$ according to Eq.~\eqref{park_pande}.
The transition to $T_2$ is attempted according to Eqs.~\eqref{st_met} and ~\eqref{st_delta}.
\item Once the trajectory at $T_2$ is sampled, accumulate potential energy and calculate $\bar{E}_{2}$.
We then, update the weight parameters $f_2$ and $f_3$.
Once the trajectory at $T_3$ is sampled, accumulate potential energy and calculate $\bar{E}_{3}$.
We then, update the weight parameters $f_3$ and $f_4$.
\item Once the trajectories at all temperatures are sampled, calculate all average potential energies, and update the weight parameters.
\end{itemize}

In addition, the following modification is made to quickly sample the entire temperature space in a system with a large average energy difference between temperatures, such as proteins in explicit solvent.
The weight parameters are calculated assuming that the average potential energy at temperatures that have never been sampled are equal to the average potential energy at the current temperature~\cite{zhang2015folding}.


\subsection{Simulated tempering for the isobaric-isothermal ensemble}
We now introduce an example of the multidimensional ST for the isobaric-isothermal ensemble~\cite{mori2010generalized}.
We refer to this method as pressure-temperature simulated tempering (PTST), which means simulated tempering in pressure and temperature space.
In PTST, temperature and pressure become dynamical variables which take discrete values $T_n\:(T_1<T_2<...<T_N)$ and $P_m\:(P_1<P_2<...<P_M)$.
Probability distribution of a state at temperature $T_n$ and potential energy $E$, pressure $P_m$ and system volume $V$ is given by the following generalised isobaric-isothermal distribution:
\begin{align}
W_\mathrm{ST}(E,V;T_n,P_m)=\mathrm{e}^{-{\beta}_n(E+P_mV)+g_{n,m}},
\end{align}
where $\beta_{n}=1/k_\mathrm{B}T_n$.
The function $g_{n,m}=g(T_n,P_m)$ is chosen so that the distribution function $p_\mathrm{ST}(T_n,P_m)$ of temperature and pressure may be uniform:
\begin{align}
p_\mathrm{ST}(T_n,P_m) &=\int\!\!\!\int dVdEn(E,V)W_\mathrm{ST}(E,V;T_n,P_m)\notag\\
&=\int\!\!\!\int dVdEn(E,V)\mathrm{e}^{-\beta_n(E+P_mV)+g_{n,m}}\notag\\
&\equiv\mathrm{const.} \label{conv_ptst}
\end{align}
where $n(E,V)$ is the density of states.
From Eq.~\eqref{conv_ptst}, the function $g_{n,m}$ is given by
\begin{equation}
\label{gibbs}
g_{nm} = -\ln\int\!\!\!\int dVdEn(E,V)\mathrm{e}^{-\beta_n(E+P_mV)}.
\end{equation}
Hence, $g_{n,m}$ is the dimensionless Gibbs free energy at temperature $T_n$ and pressure $P_m$.

Once the parameters $g_{n,m}$ are determined, a PTST simulation is realised by repeating the following two steps:
\begin{enumerate}
\item Perform an isobaric-isothermal MC or MD simulation for a certain number of steps at temperature $T_n$ and pressure $P_m$.
\item Update the temperature $T_n$ to the neighbour value $T_{n\pm 1}$ ($T$-update) or $P_m$ to the neighbour value $P_{m\pm 1}$ ($P$-update) while fixing the configurations. The transition probability of this process is given by the following Metropolis criterion:
\end{enumerate}
\begin{equation}
\label{ptst_met}
w(T_n,P_m;T_{n^{\prime}},P_{m^{\prime}})=\min(1,\exp(-\Delta)),
\end{equation}
where
\begin{equation}
\label{ptst_update}
\Delta=(\beta_{n^{\prime}}-\beta_{n})E+(\beta_{n^{\prime}}P_{m^{\prime}}-\beta_nP_m)V-(g_{n,m}-g_{n^{\prime},m^{\prime}}).
\end{equation}
For $T$-update, i.e., $n^{\prime}=n\pm1$ and $m^{\prime}=m$, Eq.~\eqref{ptst_update} reads
\begin{equation}
\label{t_update}
\Delta=\Delta_{T}=(\beta_{n\pm1}-\beta_{n})(E+P_{m}V)-(g_{n,m}-g_{n\pm1,m}).
\end{equation}
For $P$-update, i.e., $n^{\prime}=n$ and $m^{\prime}=m\pm1$, Eq.~\eqref{ptst_update} reads
\begin{equation}
\label{p_update}
\Delta=\Delta_P=\beta_n(P_{m\pm1}-P_m)V-(g_{n,m}-g_{n,m\pm1}).
\end{equation}


\subsection{On-the-fly weight determination for the isobaric-isothermal simulated tempering}

First, note that the one-dimensional on-the-fly weight determination scheme in the canonical ensemble described above is equivalent to calculating the following relation of dimensionless free energy using trapezoidal rules:
\begin{align}
f(\beta)=\int^{\beta}_{\beta_0}\bar{E}(\beta)d\beta'.
\end{align}

In the NPT ensemble, we consider the following formula:
\begin{align}
g(\beta,P)=f(\beta)+{\beta}PV,
\end{align}
where $g(\beta,P)$ is the dimensionless Gibbs free energy.

In the form of total derivative, we have
\begin{align}
dg(\beta,P)=\bar{H}(\beta,P)d{\beta}+{\beta}\bar{V}(\beta,P)dP.
\end{align}
Here, $\bar{H}(\beta,P)=\bar{E}(\beta,P)+P\bar{V}(\beta,P)$ is the average enthalpy at $\beta=1/k_\mathrm{B}T$ and pressure $P$, and $\bar{V}(\beta,P)$ is the average volume at temperature $T$ and pressure $P$.
The general solution of the above equation is written in the following form:

\begin{align}
\label{total_gibbs}
g(\beta,P)&=\int^{\beta}_{{\beta}_0}\bar{H}({\beta}',P)d{\beta}'+{\beta_0}\int^{P}_{P_0}\bar{V}(\beta_0,P')dP'\\
&=\int^{\beta}_{{\beta}_0}\bar{H}({\beta}',P_0)d{\beta}'+{\beta}\int^{P}_{P_0}\bar{V}(\beta,P')dP'.
\end{align}
In ST, $T$-update and $P$-update are independent of each other, therefore instead of $g(\beta,P)$, the following two independent weight parameters $g^T(\beta,P)$ and $g^P(\beta,P)$ can be used for each update.

For $T$-update:
\begin{align}
\label{gibbs_tupdate}
g^T(\beta,P)=\int^{\beta}_{{\beta}_0}\bar{H}({\beta}',P)d{\beta}'.
\end{align}

For $P$-update:
\begin{align}
\label{gibbs_pupdate}
g^P(\beta,P)={\beta}\int^{P}_{P_0}\bar{V}(\beta,P')dP'.
\end{align}

The dimensionless Gibbs free energy can be obtained as follows using Eq.~\eqref{total_gibbs}:

\begin{align}
g(\beta,P)&=g^T(\beta,P)+g^P(\beta_0,P)\\
&=g^P(\beta,P)+g^T(\beta,P_0).
\end{align}

The weight parameters are obtained by calculating Eq.~\eqref{gibbs_tupdate} and Eq.~\eqref{gibbs_pupdate} using the trapezoidal rule:
\begin{align}
\label{park_pande_t}
g^T_{n+1,m}=g^T_{n,m} + ({\beta}_{n+1}-{\beta}_{n})\frac{\bar{H}_{n+1,m}+\bar{H}_{n,m}}{2},
\end{align}
\begin{align}
\label{park_pande_p}
g^P_{n,m+1}=g^P_{n,m} + \beta_{n}(P_{m+1}-P_{m})\frac{\bar{V}_{n,m+1}+\bar{V}_{n,m}}{2}.
\end{align}
Here, Eq.~\eqref{park_pande_t} is the Park-Pande formula in NPT ensemble, and Eq.~\eqref{park_pande_p} is a generalisation of the Park-Pande formula to the pressure update.

Since $T$-update and $P$-update are independent, Eq.~\eqref{t_update} and Eq.~\eqref{p_update} are finally as follows given by
\begin{equation}
\label{tupdateonthe}
\Delta=\Delta_T=(\beta_{n\pm1}-\beta_{n})(E+P_{m}V)-(g^T_{n,m}-g^T_{n\pm1,m}),
\end{equation}
\begin{equation}
\label{pupdateonthe}
\Delta=\Delta_P=\beta_n(P_{m\pm1}-P_m)V-(g^P_{n,m}-g^P_{n,m\pm1}).
\end{equation}

The protocol for determining the weight parameters is as follows:

\begin{itemize}
\setlength{\parskip}{0cm}
\setlength{\itemsep}{0cm} 
\item First, set weight parameters $g^T_{n,m}=0$ and $g^P_{n,m}=0$.
\item Start the PTST simulation at the lowest temperature $T_1$ and lowest pressure $P_1$.
Accumulate potential energy and calculate the average potential energy $\bar{E}_{1,1}$, and accumulate system volume and calculate the average volume $\bar{V}_{1,1}$.
We then obtain two weight parameters $g^T_{2,1}$ and $g^P_{1,2}$ according to Eq.~\eqref{park_pande_t} and Eq.~\eqref{park_pande_p}.
The transition to $T_2$ and $P_2$ is attempted according to Eqs.~\eqref{ptst_met} and \eqref{ptst_update} with Eq.~\eqref{tupdateonthe} and  Eq.~\eqref{pupdateonthe}, respectively.
\item Once the trajectory at $T_2$ is sampled, accumulate potential energy and calculate $\bar{E}_{2,1}$.
Also accumulate system volume and calculate $\bar{V}_{2,1}$.
We then, update all weight parameters at ($T_2$, $P_1$) and the nearest temperatures and pressures.
Once the trajectory at $P_2$ is sampled, accumulate potential energy and calculate $\bar{E}_{1,2}$.
Also accumulate system volume and calculate $\bar{V}_{1,2}$.
We then, update all weight parameters at ($T_1$, $P_2$) and the nearest temperatures and pressures.
\item Once the trajectories at all temperatures and pressures are sampled, calculate all average potential energies and average system volumes, and update the weight parameters.
\end{itemize}

In addition, the following modification is made to quickly sample the entire PT space in a system with a large average energy difference between temperatures such as proteins in explicit solvent.
The weight parameters are calculated assuming that the average potential energy and average volume at temperatures and pressures that have never been sampled are equal to the average potential energy and average volume at the current temperature and pressure~\cite{zhang2015folding}.



\section{Computational details}
\label{sec:computational}
We performed a MD simulation using the PTST method with the fast on-the-fly weight determination. 
We used a small peptide, chignolin, in explicit water as the simulation system.
The system consists of chignolin, 923 water molecules, and two sodium ions.
The total number of atoms in the system was 2,909.
The system was placed in a cubic unit cell with periodic boundary conditions.

The MD simulation was carried out by the NAMD program package (version 2.13)~\cite{phillips2005scalable}.
We implemented a Tcl script that performs the PTST with on-the-fly weight determination.
The CHARMM22 force field~\cite{mackerell1998all} with the CMAP corrections~\cite{mackerell2004extending} was used for chignolin, and the TIP3P~\cite{jorgensen1983comparison} model was used for the water molecules.
The temperature was controlled by the Langevin thermostat.
While there exist several methods for barostat such as those in Refs.~\cite{klein1,klein2}, we controlled the pressure by the  Nos\'{e}-Hoover Langevin piston barostat~\cite{phillips2005scalable,quigley2004langevin}.
The electrostatic interactions were calculated using the particle mesh Ewald method (PME)~\cite{darden1993particle,essmann1995smooth}.
The cutoff distance for the van der Waals interactions were set to $12.0$ \AA .
The SETTLE algorithm was used to constrain the vibration of all bonds involving hydrogen atoms~\cite{Miyamotosettle}.
The time step was set to 2.0 fs.

\subsection{Algorithm test on small P-T space and comparison between PTST and PTREMD}
We used the following five temperature ($T_1,...,T_5$) and four pressure ($P_1,...,P_4$) values: 300.0, 308.2, 316.6, 325.0, and 334.0 K for temperature and 0.1, 32.5, 65.0, and 100.0 MPa for pressure.
Trials of ST update were performed every $1.0$ ps and the trajectory data were stored just before the trials.
At the ST trial, either updating temperature or updating pressure was chosen randomly and then either $T_{n-1}$ or $T_{n+1}$ and $P_{m-1}$ or $P_{m+1}$ for each update was also chosen randomly.
The total simulation time was $1.0$ $\rm{\mu s}$.

We have performed an isobaric-isothermal replica-exchange molecular dynamics (PTREMD) simulation~\cite{okabe2001replica,sugitaokamoto2002} with the same parameter values for $T_n$ and $P_m$ as above (the total number of replicas is then $5\times 4=20$).
The replica exchange were tried every $1.0$ ps and the trajectory data were stored just before the trials.
At the replica-exchange trial, either exchanging temperature or exchanging pressure was chosen randomly and then either pairs of $\{(T_1,T_2),(T_3,T_4)\}$ or $\{(T_2,T_3),(T_4,T_5)\}$ and pairs of $\{(P_1,P_2),(P_3,P_4)\}$ or $\{(P_2,P_3),(P_1,P_4)\}$ for each exchange was also chosen randomly.
Each replica was simulated for $80$ ns.
The total simulation time was $1.6$ $\rm{\mu s}$.
As a reference of the weight parameters, we calculated dimensionless Gibbs free energy using the MBAR method~\cite{shirts2008statistically} for the potential energy and system volume obtained from this PTREMD simulation.

\subsection{Test in practical space sizes and structural sampling of chignolin under low temperature and high pressure}
We used the following 16 temperature ($T_1,...,T_{16}$) and 15 pressure ($P_1,...,P_{15}$) values: 300.0, 308.2, 316.6, 325.0, 334.0, 343.2, 352.5, 362.1, 372.0, 382.1, 392.5, 403.2, 414.2, 425.5, 437.1, and 450.0 K for temperature and 0.1, 30.0, 60.0, 90.0, 120.0, 150.0, 180.0, 210.0, 240.0, 270.0, 300.0, 330.0, 360.0, 390.0 and 420.0 MPa for pressure.
Trials of ST update were performed every $1.0$ ps and the trajectory data were stored just before the trials.
At the ST trial, either updating temperature or updating pressure was chosen randomly and then either $T_{n-1}$ or $T_{n+1}$ and $P_{m-1}$ or $P_{m+1}$ for each update was also chosen randomly.
Five ST simulations of $1.6$ $\rm{\mu s}$ with different initial conditions were performed.
The total simulation time was $8.0$ $\rm{\mu s}$.
The total of $7.5$ $\rm{\mu s}$ was used for analysis, excluding the first $100$ ns during weight parameter determination for each ST simulation. 


\section{Results}
\label{sec:result}
\subsection{Algorithm test on small P-T space and comparison between PTST and PTREMD}
\begin{figure}[htb]
\begin{center}
\includegraphics[scale=0.35]{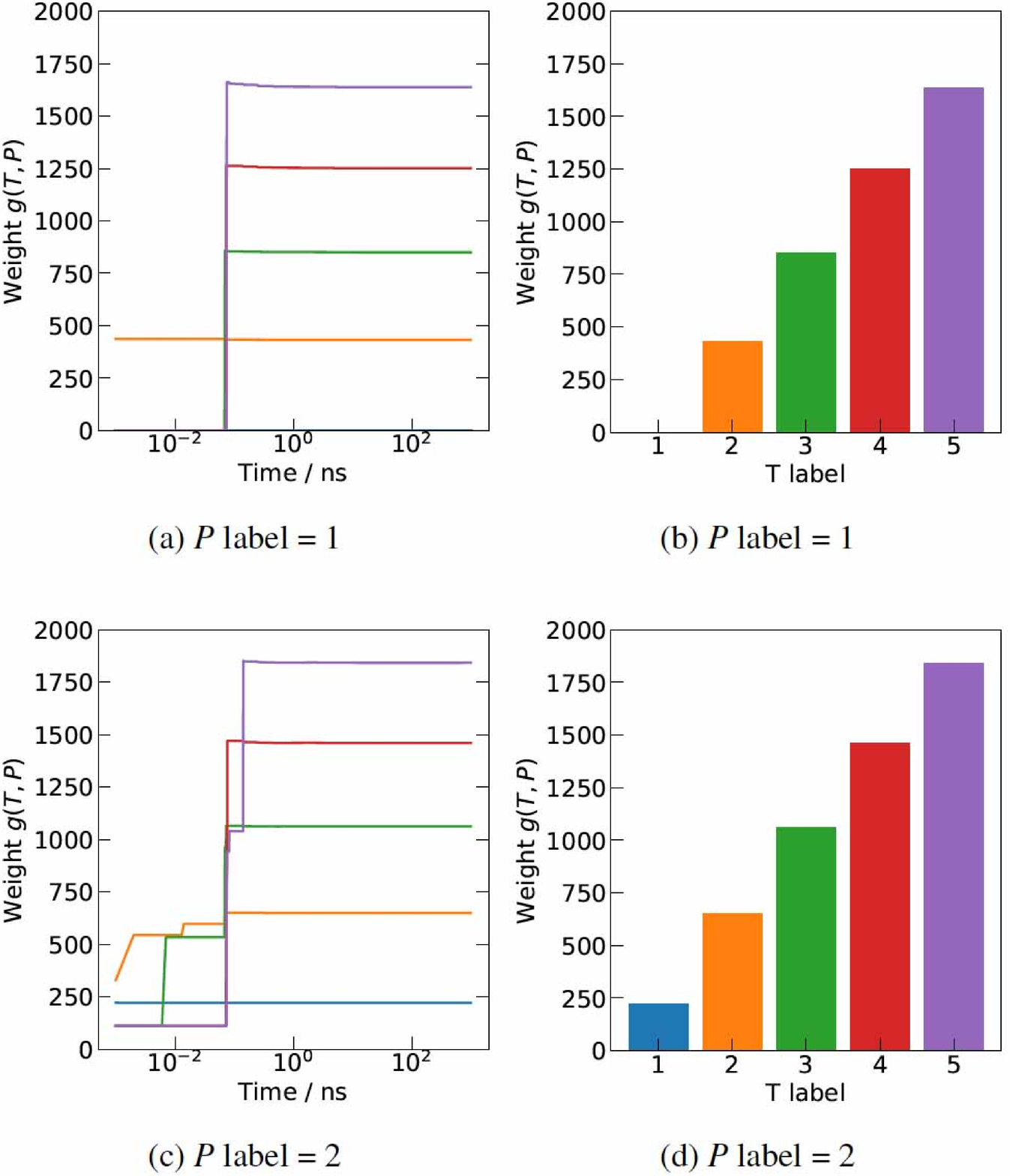}
\end{center}
\caption{\label{index20_gibbs} PTST simulation with on-the-fly weight determination. (a), (c) Time evolution of the weight parameters during the PTST simulation. (b), (d) Weight parameters obtained from the 1.6 $\mu s$ PTREMD simulation. Each color in the left time series corresponds to the temperature label in the right bar graph.}
\end{figure}

The time evolution of the weight parameters at two selected pressure labels is shown in 
Fig.~\ref{index20_gibbs}(a) and Fig.~\ref{index20_gibbs}(c).
We normalised the weight parameters with $g(T_1,P_1)=0$.
As comparison, the weight parameters obtained from the 1.6 $\rm{\mu s}$ the PTREMD simulation 
under the same conditions are shown in 
Fig.~\ref{index20_gibbs}(b) and Fig.~\ref{index20_gibbs}(d).
Immediately after starting the simulation with all weight parameters set to zero, the values were updated to non-zero values.
By 1.0 ns, the weight parameters reached plateau values and converged to the values obtained from the 1.6 $\rm{\mu s}$ PTREMD simulation.

\begin{figure}[htb]
\begin{center}
\includegraphics[scale=0.5]{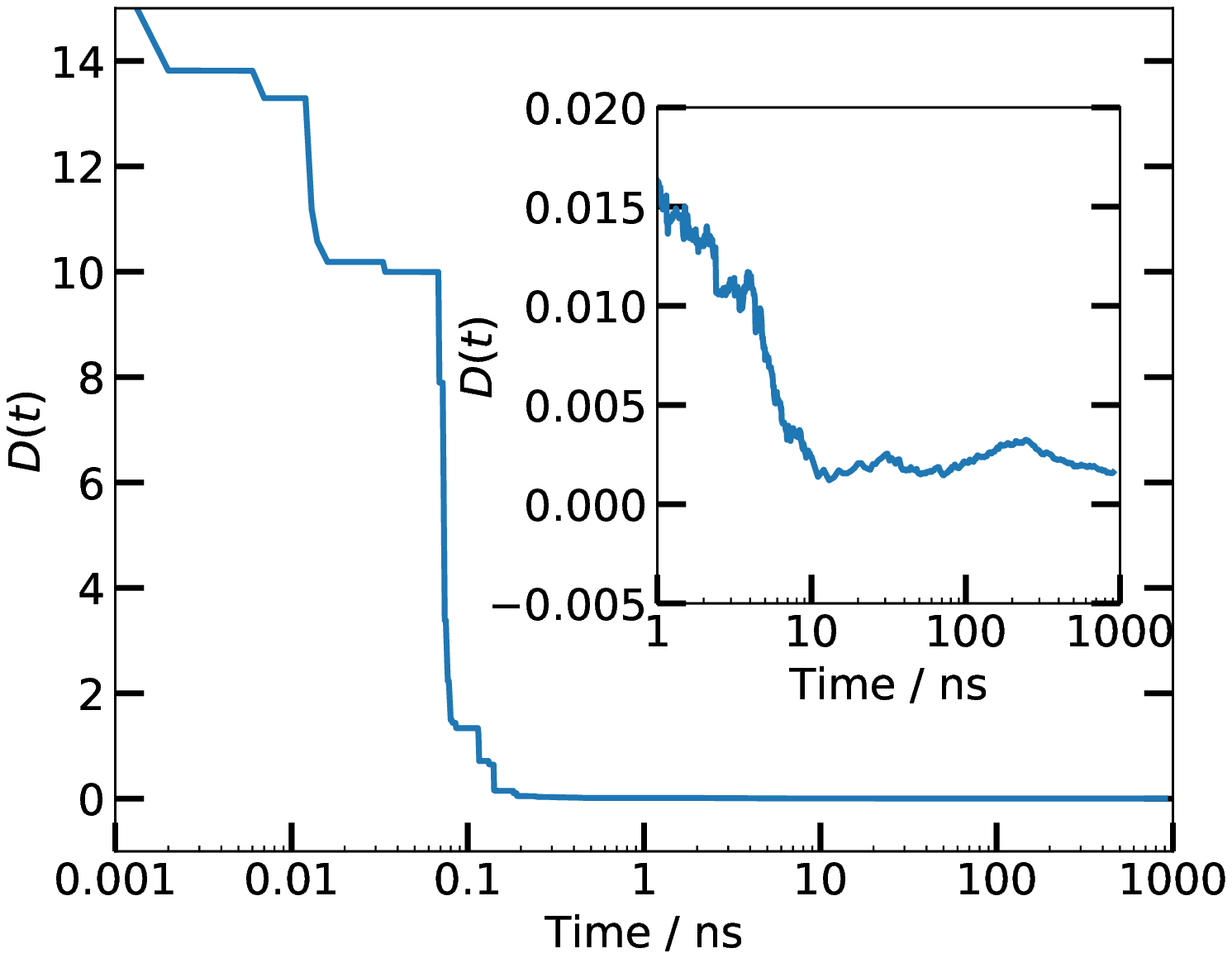}
\end{center}
\caption{\label{diffseries_par} Time evolution of $D(t)$ during PTST simulation and enlarged graph of this time series after 1 ns.}
\end{figure}

In order to investigate the convergence accuracy of weight parameters, the difference between the weight parameters obtained from the 1.6$\rm{\mu s}$ PTREMD simulation and the weight parameters calculated during the PTST simulation was evaluated using the following formula:
\begin{equation}
D(t)=\sum_i \biggl|\frac{g^\mathrm{ST}_i(t)-g^\mathrm{REMD}_i}{g^\mathrm{REMD}_i}\biggl|.
\end{equation}
Where, $g^\mathrm{ST}_i(t)$ is the weight parameter of parameter label $i$ obtained at time $t$ during the PTST simulation, and $g^\mathrm{REMD}_i$ is the weight parameter of parameter label $i$ obtained by the PTREMD simulation.
The parameter label $i$ corresponds to the combination of temperature and pressure.
The time evolution of $D(t)$ is shown in Fig.~\ref{diffseries_par}.
$ D (t) $ approached 0 rapidly around 0.1 ns and almost reached 0 at 10 ns.
Therefore, it can be seen that weight parameters converged to weight parameters obtained from the 1.6 $\rm{\mu s}$ PTREMD simulation after 10 ns of the PTST simulation.

\begin{figure*}[htb]
\begin{center}
\includegraphics[scale=0.4]{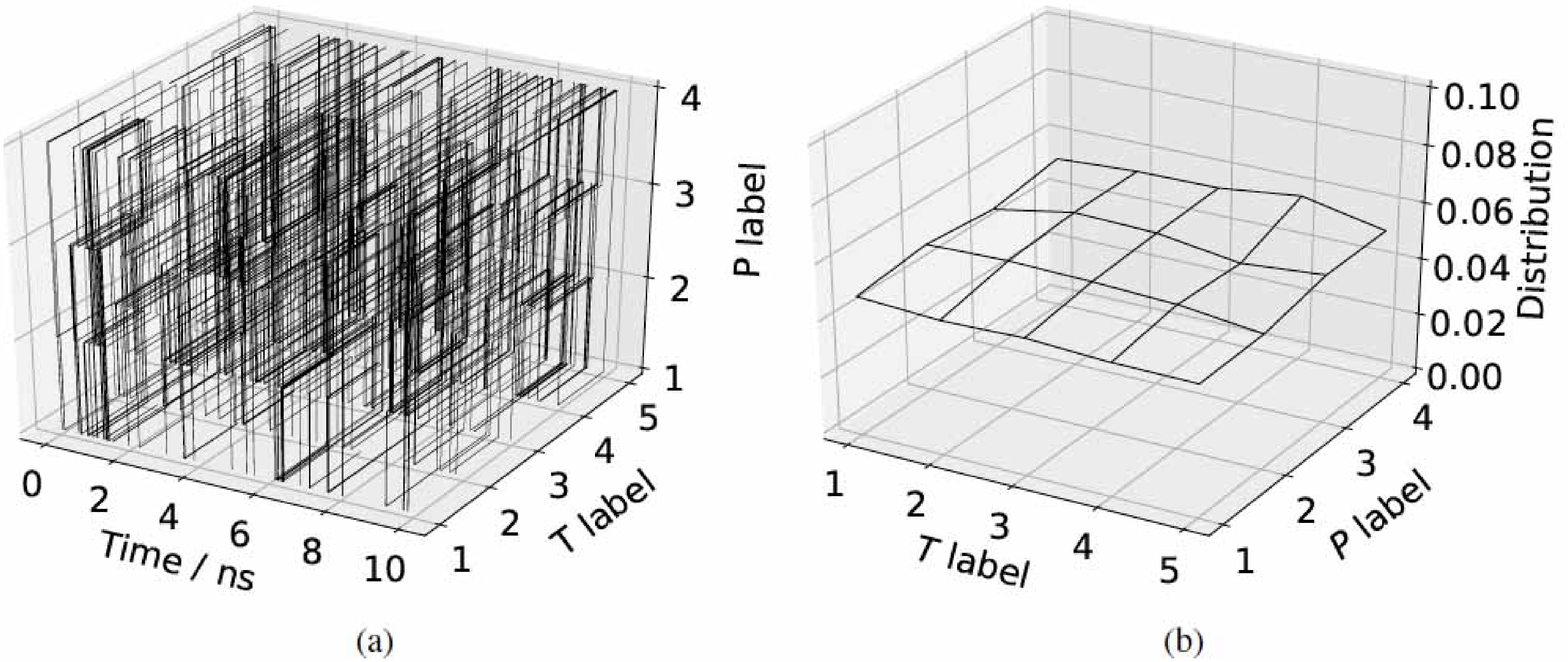}
\end{center}
\caption{\label{index20pt}Results of the PTST simulation with on-the-fly weight determination. (a) First $10$ ns time series of $T$ and $P$ and (b) distribution of $T$ and $P$.}
\end{figure*}

The time series of temperature and pressure labels is shown in 
Fig.~\ref{index20pt}(a).
This Figure shows that the random walk in the entire parameter space was sufficiently realised in this simulation.
The histogram of temperature and pressure labels is shown in 
Fig.~\ref{index20pt}(b).
A flat histogram expected by Eq.~\eqref{conv_ptst} was obtained.
Hence, the PTST simulation was appropriately performed by our proposed method.

As in the comparative studies between ST and REM~\cite{zhang2008comparison,park2008comparison}, we compared the transition probabilities between PTST and PTREMD.
In this study, we define the transition ratio $P_\mathrm{TR}(i,j)$ as follows for comparison:

\begin{equation}
P_\mathrm{TR}(i,j)=\frac{n_{i,j}}{N_{i,j}},
\end{equation}
where $n_{i,j}$ is the total acceptance count of transition from temperature label $i$ and pressure label $j$ to neighbour labels, $N_{i,j}$ is the total number of trial of transitions from temperature label $i$ and pressure label $j$ to neighbour labels.
The transition ratio $P_\mathrm{TR}(i,j)$ is listed in Table~\ref{tab:transition}.
The results show that PTST performed under the same conditions has the transition ratio about twice that of PTREMD.

\begin{table*}[htb]
\caption{\label{tab:transition}The transition ratio $P_\mathrm{TR}(i,j)$ of PTST and PTREMD simulations.}
\begin{ruledtabular}
\begin{tabular}{ccccccccc}
  &\multicolumn{2}{c}{$P_1$}&\multicolumn{2}{c}{$P_2$}&\multicolumn{2}{c}{$P_3$}&\multicolumn{2}{c}{$P_4$}\\
 &PTST&PTREMD&PTST&PTREMD&PTST&PTREMD&PTST&PTREMD\\ \hline
$T_1$& 0.17 & 0.08 & 0.26 & 0.14 & 0.26 & 0.14 & 0.17 & 0.08\\
$T_2$& 0.24 & 0.13 & 0.34 & 0.17 & 0.34 & 0.18 & 0.24 & 0.14\\
$T_3$& 0.25 & 0.12 & 0.34 & 0.18 & 0.34 & 0.18 & 0.25 & 0.13\\
$T_4$& 0.25 & 0.12 & 0.33 & 0.17 & 0.34 & 0.18 & 0.24 & 0.12\\
$T_5$& 0.16 & 0.08 & 0.26 & 0.13 & 0.26 & 0.14 & 0.17 & 0.09
\end{tabular}
\end{ruledtabular}
\end{table*}

\subsection{Test in practical space sizes and structural sampling of chignolin under low temperature and high pressure}

\begin{figure*}[htb]
\begin{center}
\includegraphics[scale=0.35]{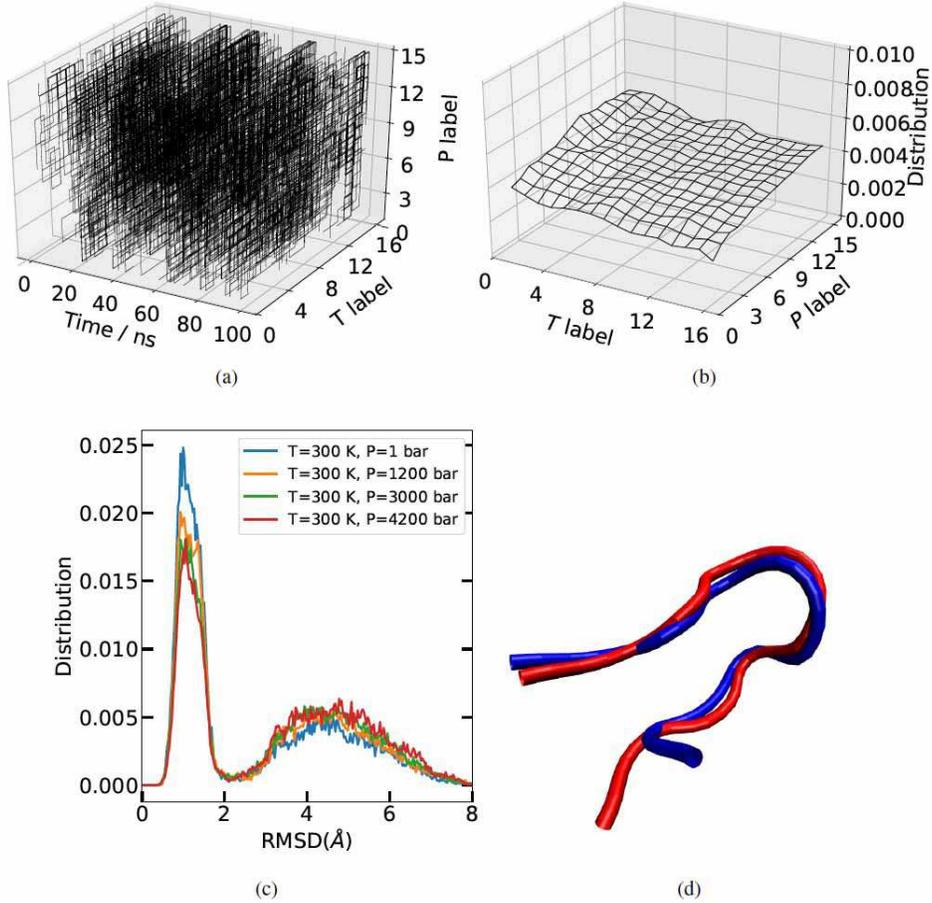}
\end{center}
\caption{\label{160walkfel}Results of the PTST simulation with on-the-fly weights determination. (a) The last $100$ ns time series of $T$ and $P$. (b) Distribution of $T$ and $P$. (c) Distribution of RMSD at $T=300$ K. (d) Backbone structures of chignolin obtained by the ST simulation at $T=300$ K, $P=1$ bar (in red) and the NMR structure (in blue). }
\end{figure*}

The time series of temperature and pressure labels is shown in 
Fig.~\ref{160walkfel}(a).
This Figure shows that the random walk in the entire parameter space was sufficiently realised in this simulation.
The histogram of temperature and pressure labels is shown in 
Fig.~\ref{160walkfel}(b).
This histogram is not perfectly uniform.
However, there are no excessively sampled temperatures and pressures and no unsampled temperatures and pressures. 
The distribution of the root-mean-square distance (RMSD) at $T=300$ K is shown in 
Fig.~\ref{160walkfel}(c).
We used the NMR structure of chignolin (PDB ID: 1UAO, Model 1) as the reference structure for the RMSD calculations.
Here, RMSD was obtained with respect to $\mathrm{C_\alpha}$, $\mathrm{C}$ and $\mathrm{N}$ atoms in the backbone.
The figure shows that the simulation by our proposed method escapes from the trapped structure and samples the conformational space extensively.
Fig.~\ref{160walkfel}(d) shows the structure of the left peak in 
Fig.~\ref{160walkfel}(c).
The proportion of $\beta$-hairpin structure (left peak in 
Fig.~\ref{160walkfel}(c)) that is stable at $P = 1$ bar 
decreases depending on pressure, and at $P = 4200$ bar, unfolded structures are dominant.

From the above results, our method was able to realise a random walk for the P-T space size that was practically used and was able to perform conformational sampling of proteins under low and high pressures.

\section{Conclusions}
\label{sec:conclusion}
In this article, we proposed a method to extend the fast on-the-fly weight determination scheme for simulated tempering to 
a two-dimensional space of temperature and pressure.
This method considers independent two weight parameters in each update of temperature and pressure of two-dimensional simulated tempering and self-updates the weight parameters during the PTST simulation by calculating the thermodynamic relations using the trapezoidal rule.
The algorithm was tested using chignolin in explicit water.
During the simulated tempering simulation, weight parameters were self-updated and converged rapidly to weight parameters obtained from a long REMD simulation, and a uniform random walk in two-dimensional space of temperature and pressure was realised.
When using REMD to estimate the weight parameters of two-dimensional simulated tempering, a very large number of CPUs are required.
In this method, it is not necessary to perform REMD in advance.
Therefore, simulation for structure sampling and determination of weight parameters can be performed with fewer computational resources.
In addition to the above, the comparison of the transition ratio with two-dimensional REMD simulation showed that simulated tempering has higher transition ratio than replica exchange even in the two-dimensional space.

With our method, it can be said that simulated tempering has become a practical and easy-to-use tool for a wider range of problems.
Its applications to high-pressure denaturation of larger proteins than small peptides are now in progress.

\section*{Acknowledgement}
All the computer simulations in this research were performed using supercomputers at Research Center for Computational Science, Okazaki, Japan.

%
\noindent

\end{document}